\documentclass[prl,aps,twocolumn]{revtex4}
\usepackage{graphicx}

\newcommand{\be}{\begin{equation}}
\newcommand{\ee}{\end{equation}}
\newcommand{\ba}{\begin{eqnarray}}
\newcommand{\ea}{\end{eqnarray}}
\newcommand{\ban}{\begin{eqnarray*}}
\newcommand{\ean}{\end{eqnarray*}}

\newcommand{\ket}[1]{\mbox{$ \left| #1 \right\rangle $}}

\newcommand{\demi}{\frac{1}{2}}

\newcommand{\one}{\leavevmode\hbox{\small1\normalsize\kern-.33em1}}

\begin{document}

\title{Fast and simple one-way Quantum Key Distribution}
\author{Damien Stucki, Nicolas Brunner, Nicolas Gisin, Valerio Scarani, Hugo Zbinden}
\address{Group of Applied Physics, University of Geneva, 20, rue de
l'Ecole-de-M\'edecine, CH-1211 Geneva 4, Switzerland}

\date{\today}

\begin{abstract}

We present and demonstrate a new protocol for practical quantum
cryptography, tailored for an implementation with weak coherent
pulses to obtain a high key generation rate. The key is obtained
by a simple time-of-arrival measurement on the {\em data line};
the presence of an eavesdropper is checked by an interferometer on
an additional {\em monitoring line}. The setup is experimentally
simple; moreover, it is tolerant to reduced interference
visibility and to photon number splitting attacks, thus featuring
a high efficiency in terms of distilled secret bit per qubit.
\end{abstract}

\maketitle

Quantum key distribution (QKD) is the only method to distribute a
secret key between two distant authorized partners, Alice and Bob,
whose security is based on the laws of physics \cite{review}. QKD
is the most mature field in quantum information; nevertheless,
there is still some work ahead in order to build a practical
system which is reliable and at a same time fast and provably
secure. This paper presents an important improvement in this
direction. The quest for {\em rapidity} is the inspiring
motivation of this system: the idea is to obtain the secret bits
from the simplest possible measurement (here, the time of arrival
of a pulse) without introducing lossy optical elements at Bob's.
{\em Security} is obtained by occasionally checking quantum
coherence: in QKD, a decrease of coherence is attributed to the
presence of the eavesdropper Eve, who has attacked the line and
obtained some information on the bit values, at the price of
introducing errors. {\em Reliability} is achieved by using
standard telecom components; in particular, the source is an
attenuated laser, and bits are encoded in time-bins, robust
against polarization effects in fibers. In this paper, we first
define the protocol and demonstrate its advantages: simplicity,
and robustness against both reduced interference visibility and
photon number splitting (PNS) attacks \cite{pns}. Then, we present
a first proof-of-principle experiment.

\begin{figure}
\includegraphics[width=8cm]{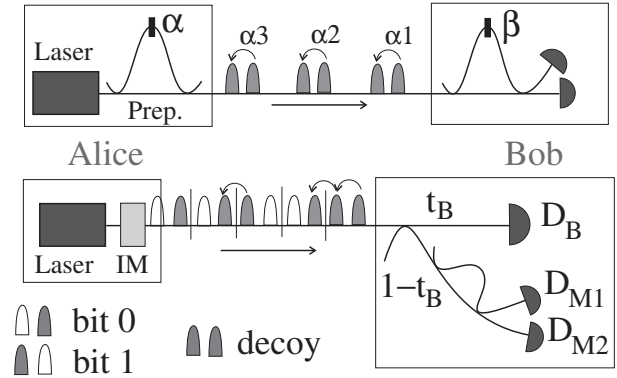} \caption{Comparison of the currently implemented BB84 protocol
with phase encoding (Top) with the scheme proposed here (Bottom).
Arrows over pulses indicate coherence (phase set to 0 in our
scheme). See text for details.} \label{figproto}
\end{figure}

{\em The protocol.} To date, the most developed setups for
practical QKD implement the Bennett-Brassard 1984 (BB84) protocol
\cite{bb84} using phase encoding between two time-bins, as
sketched in Fig.~\ref{figproto} Top (see \cite{review} for a
detailed description). The four states belonging to two mutually
orthogonal bases are the
$\ket{1}\ket{0}+e^{i\alpha}\ket{0}\ket{1}$ where $\alpha=0,\pi$
(bits 0 and 1 in the X basis) or
$\alpha=\frac{\pi}{2},\frac{3\pi}{2}$ (bits 0 and 1 in the Y
basis). Bob detects in the X (Y) basis by setting $\beta=0$
($\beta=\frac{\pi}{2}$). Both bases correspond thus to an
interferometric measurement. As a first step towards simplicity,
we replace (say) the Y basis with the Z basis $\{\ket{1}\ket{0},
\ket{0}\ket{1}\}$. Measuring in this basis amounts simply to the
measurement of a time of arrival, and is thus insensitive to
optical errors \cite{thales}. Bits are encoded in the Z basis,
which can be used most of the times, the X basis being used only
occasionally to check coherence \cite{unknown}.

In a practical QKD setup, the source is an attenuated laser: here,
Alice's source consists of a cw laser followed by an intensity
modulator (IM), which either prepares a pulse of mean
photon-number $\mu$ or blocks completely the beam (empty or
"vacuum" pulses) \cite{modelocked}. The $k$-th logical bit is
encoded in the two-pulse sequences consisting of a non-empty and
an empty pulse:
\ba \ket{0_k}&=&\ket{\sqrt{\mu}}_{2k-1}\ket{0}_{2k}\,,\\
\ket{1_k}&= &\ket{0}_{2k-1}\ket{\sqrt{\mu}}_{2k}\,.\ea Note that
$\ket{0_k}$ and $\ket{1_k}$ are not orthogonal, due to their
vacuum component; however, a time-of-arrival measurement, whenever
conclusive, provides the optimal unambiguous determination of the
bit value \cite{draft}. To check coherence, we produce a fraction
$f\ll 1$ of {\em decoy sequences} $\ket{\sqrt{\mu}}_{2k-1}
\ket{\sqrt{\mu}}_{2k}$; while for BB84, one should produce the two
states $\ket{\sqrt{\mu/2}}_{2k-1} \ket{\pm\sqrt{\mu/2}}_{2k}$.
Now, due to the coherence of the laser, there is a well-defined
phase between any two non-empty pulses: within each decoy
sequence, but also {\em across the bit-separation} in the case
where bit number $k$ is 1 and bit number $k+1$ is 0 (a "1-0 bit
sequence"). Since we produce equally-spaced pulses, the coherence
of both decoy and 1-0 bit sequences can be checked with a single
interferometer (see Fig.~\ref{figproto}, Bottom). And there is a
further benefit: coherence being distributed both within and
across the bit separations, Eve cannot count the number of photons
in any finite number of pulses without introducing errors
\cite{draft}: in our scheme the PNS attacks can be detected
\cite{inoue}. To detect PNS attacks in BB84, one needs to
complicate the protocol by the technique of decoy states, which
consists in varying $\mu$ \cite{decoy}.

The pulses propagate to Bob on a quantum channel characterized by
a transmission $t$, and are split at a non-equilibrated
beam-splitter with transmission coefficient $t_B\lesssim 1$. The
pulses that are transmitted ({\em data line}) are used to
establish the raw key by measuring the arrival times of the
photons. The counting rate is $R=1-e^{-\mu t  t_B \eta} \approx
\mu t t_B \eta $, where $\eta$ is the quantum efficiency of the
photon counter. The pulses that are reflected at Bob's
beam-splitter go to the interferometer that is used to check
quantum coherence ({\em monitoring line}). Indeed, when both
pulses $j$ and $j+1$ are non-empty, then only detector $D_{M1}$
can fire at time $j+1$. Coherence can be quantified by Alice and
Bob through the visibility of the interference \ba V&=&
\frac{p(D_{M1})-p({ D_{M2}})}{p(D_{M1})+p({D_{M2}})} \ea where
$p(D_{Mj})$ is the probability that detector $D_{Mj}$ fired at a
time where only $D_{M1}$ should have fired. These probabilities
are small, the average detection rate on the monitoring line being
$\demi\mu\,t\,(1-t_B)\eta$ per pulse. Still, if the bit rate is
high, meaningful estimates can be done in a reasonable time.

Let's summarize the protocol:
\begin{enumerate} \item Alice sends a large number of times "bit
0" with probability $\frac{1-f}{2}$, "bit 1" with probability
$\frac{1-f}{2}$ and the decoy sequence with probability $f$. \item
At the end of the exchange, Bob reveals for which bits he obtained
detections in the data line and when detector $D_{2M}$ has fired.
\item Alice tells Bob which bits he has to remove from his raw
key, since they are due to detections of decoy sequences
(sifting). \item Analyzing the detections in $D_{2M}$ Alice
estimates the break of coherence through the visibilities
$V_{1-0}$ and $V_{d}$ associated respectively to 1-0 bit sequences
and to decoy sequences, and computes Eve's information. \item
Finally, Alice and Bob run error correction and privacy
amplification and end up with a secret key.
\end{enumerate}

{\em Estimate of the secret key rate.} The performance of a QKD
protocol is quantified by the achievable secret key rate $R_{sk}$.
To compute this quantity, we need to introduce several parameters.
The fraction of bits kept after sifting (sifted key rate) is
$R_{s}(\mu)=[R+2p_d(1-R)]\,p_{s}$ with $R=\mu t t_B\eta$ the
counting rate due to photons defined above, $p_d$ the probability
of a dark count, and $p_{s}=1-f$ here. The amount of errors in the
sifted key is called quantum bit error rate (QBER, $Q$). Moreover,
this key is not secret: Eve knows a fraction $I_{Eve}$ of it. Some
classical postprocessing (error correction and privacy
amplification) allows to extract a key which is errorless and
secret, while removing a fraction $h(Q)+I_{Eve}$, where $h$ is
binary entropy. Thence, \ba R_{sk}&=&R_{s}(\mu)\,\big(1-h(Q)-
I_{Eve}\big)\,.\label{rsknew}\ea With this figure of merit, we can
compare our scheme to BB84 implemented using the interferometric
bases X and Y, as it is done today, with an asymmetric use of the
bases such that $p_{s}=1-f$ (BB84$_{XY}$). We require that all the
visibilities are equal: $V_X=V_Y$ in BB84$_{XY}$, $V_{1-0}=V_d$ in
our scheme --- otherwise, Alice and Bob abort the protocol. Under
this assumption, the QBER of BB84 is $Q(\mu)=[R\frac{1-V}{2}
+(1-R)p_d]p_s/R_s\, \equiv\,Q_{opt}+Q_{det}$; while in our scheme
$Q(\mu)= Q_{det}$, independent of $V$.

\begin{figure}
\includegraphics[width=8.5cm]{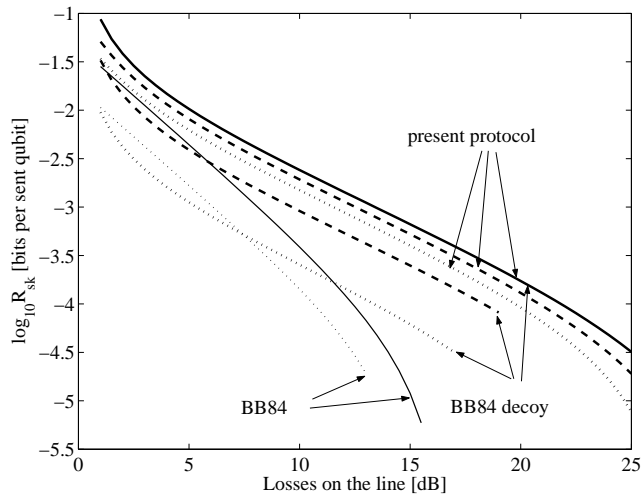} \caption{Estimate of the secret key
rates Eq.~(\ref{rsknew}) for the present protocol and for
BB84$_{XY}$ with and without decoy states, as a function of the
losses on the line $\ell$ ($t=10^{-\ell/10}$). Parameters:
$\eta=10\%$, $p_d=10^{-5}$, $t_B=1$ and $f=0.1$. Visibility: $V=1$
(full lines, identical for the two first protocols), $V=0.9$
(dashed lines) and $V=0.8$ (dotted lines; $R_{sk}=0$ for BB84
without decoy states).} \label{figcompare}
\end{figure}

In order to estimate $I_{Eve}$, we restrict the class of Eve's
attacks \cite{draft}, waiting for a full security analysis.
Because of losses and the existence of multi-photon pulses, Eve
can gain full information on a fraction of the bits without
introducing any errors. This fraction is either $r=\mu(1-t)$ or
$r=\frac{\mu}{2t}$, according to whether PNS attacks don't or do
introduce errors \cite{draft,pns}. Then Eve performs the
intercept-resend attack on a fraction $p_{IR}$ of the remaining
pulses. In BB84$_{XY}$, she introduces the error
$(1-r)p_{IR}\frac{1}{4}= \frac{1-V}{2}$ and gains the information
$I=(1-r)p_{IR}\demi=1-V$. On the present protocol, the IR will be
performed in the time basis, so $I=(1-r)p_{IR}$. However, since we
use only one decoy sequence, if Eve detects a photon in two
successive pulses she knows what sequence to prepare; the
introduced error is thence $1-V=I\xi$ with $\xi=\frac{2e^{-\mu
t}}{1+e^{-\mu t}}$ the probability that Eve detects something in
one pulse and nothing in the other. Plugging $Q(\mu)$ and
$I_{Eve}=r+I$ into Eq.~(\ref{rsknew}), we have $R_{sk}$ as an
explicit function of $\mu$; Alice and Bob must choose $\mu$ in
order to maximize it. The result of numerical optimization is
shown in Fig.~\ref{figcompare} \cite{noteopt}. As expected, the
present protocol is more robust than BB84$_{XY}$ against the
decrease of visibility.

\begin{figure}[tbp]
\includegraphics[width=8cm]{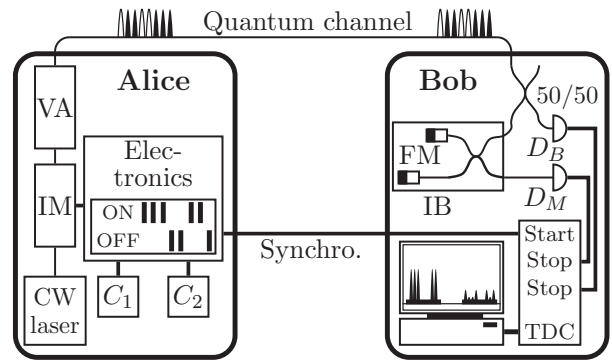}
\caption{Experimental setup (see text for details).}
\label{Fig:simpExpScheme}
\end{figure}

{\em Proof-of-principle experiment.} We show that a reasonably low
QBER and good visibility can be obtained using standard telecom
components in an implementation with optical fibers. The
experimental setup is sketched in Fig.~\ref{Fig:simpExpScheme}.
The light of a cw laser (wavelength 1550 nm) passes through a
intensity modulator (IM), which prepares the chosen pulse
sequence. For simplicity, we send always the same 8-pulse sequence
as shown in the figure, namely the string $D010$, where $D$ stands
for a decoy sequence. The frequency of 434 MHz of clock $C_{1}$
defines the time $\tau$ between two successive pulses. The
frequency of logical bits in a sequence is half of this frequency.
The clock $C_{2}$ at 600 kHz generates the delay between two
successive sequences. After the modulator, the light is attenuated
by the variable attenuator (VA) in order to obtain $\mu = 0.5$ for
5 dB loss in the quantum channel \cite{noteopt}. The
synchronization signal directly starts the time-to-digital
converter (TDC) and triggers the detectors on Bob's side. The
detectors $D_B$ (data line) and $D_M$ (monitoring line) are opened
with gates of 25 ns accepting the whole sequence, featuring
quantum efficiency $\eta=10\%$ and a dark count probability
$p_{d}=2.5\times 10^{-5}$ per ns. Of course, due to the deadtime
of the detectors, only one event per sequence and detector can be
detected. The stop signal from $D_M$ arm is delayed, which allows
to record the events of both detectors by the same TDC. The
Michelson interferometer of the monitoring line has the same path
length difference $\tau$ (46 cm of optical fibre) corresponding to
the clock frequency. It is enclosed in an insulated, temperature
controlled box (IB). The phase can be changed by changing the
temperature. The interferometer (hence our entire setup) is
polarization insensitive due to Faraday mirrors (FM) and features
a classical fringe visibility of 99\%.

The raw detection rate is of 17.0$\pm$0.1 kHz. The detection rate
is limited by the detectors, due to the 10 $\mu$s deadtime we have
to introduce in order to limit afterpulses. With current
detectors, the potential of an improved setup continuously sending
pulses at $C_{1}$, with optimized values for $\mu$, $f$ and
$t_{B}$ could only be exploited at long distances. Otherwise, one
could use a detection system based on up-conversion and fast thin
silicon detector \cite{thew05}.

\begin{figure}[tbp]
\includegraphics[width=8cm]{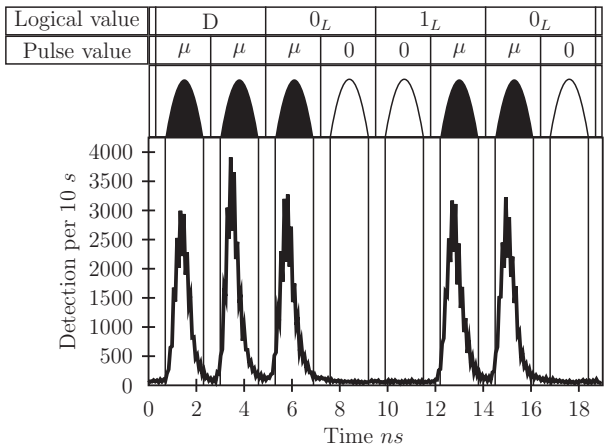}
\caption{Detection as a function of the difference of time between
start and detection. Logical values and pulse values are depicted
in more of the measurement. The difference of amplitude of the
different peaks is due to variation in  efficiency in the
detection gate.} \label{FIG:keyGate}
\end{figure}

The QBER for the pulse sequences '10' and '01' is obtained by
considering the time windows of 1.7 ns as indicated in
Fig.~\ref{FIG:keyGate}. The value is $Q=5.2\pm 0.4\%$. The
contribution of detector noise and afterpulses (which are rather
high for the long gates and high repetition rates we are using) is
estimated to be 4\%; we attribute the remaining 1\% to unperfect
intensity modulation, mainly due to too slow electronics and to
the jitter of the detectors.

The visibility of the interfering pulses on detector $D_{M}$ is
measured by varying the phase (i.e the temperature) of the
interferometer. The raw visibility is $V_{raw}\geq 92\%$, if we
consider 1.7 ns time windows. The net visibility, obtained
deducing the dark counts and afterpulses is $V\approx 98\%$. We
attribute the slight reduction of the visibility to a non perfect
overlap of the interfering pulses due to timing jitter and
fluctuations in the intensity modulation. However, this reduced
visibility has no significant consequence on the secret key rate
(Fig.~\ref{figcompare}). This tolerance in visibility simplifies
the adjustment of the interferometers. With our basic thermal
stabilization the interferometer needed to be readjusted only
about every 30 minutes. Indeed, for our path length difference, a
temperature stability of 0.01 K guarantees $V\gtrsim 80\%$. Note,
the higher $C_{1}$, the easier becomes the stabilization of the
interferometer.

{\em Conclusion.} We have introduced a new scheme for QKD and
presented first experimental results. The scheme features several
advantages: The data line is very simple, with low losses at Bob's
side and small optical QBER. The scheme is tolerant against
reduced interference visibility and is robust against PNS attacks
(thus allowing the mean photon number to be large, typically
$\mu\approx 0.5$). Finally, it is polarization insensitive. The
existence of such a scheme shows that the main limiting parameter
for practical quantum cryptography are the imperfections of the
detectors.

We acknowledge financial support from the Swiss NCCR "Quantum
photonics" and the European Project SECOQC, and thank Avanex for
the loan of an intensity modulator.


\begin{thebibliography}{10}



\bibitem{review} N. Gisin, G. Ribordy, W. Tittel and H.
Zbinden, Rev. Mod. Phys. {\bf 74}, 145 (2002)

\bibitem{pns} G. Brassard, N. L\"{u}tkenhaus, T. Mor, B.C. Sanders, Phys. Rev. Lett. {\bf 85}, 1330
(2000); N. L\"{u}tkenhaus, Phys. Rev. A {\bf 61}, 052304 (2000)

\bibitem{bb84} C.H. Bennett, G. Brassard, in:
Proceedings IEEE Int. Conf. on Computers, Systems and Signal
Processing, Bangalore, India (IEEE, New York, 1984), pp. 175-179.

\bibitem{thales} Our data line is that of a classical communication
channel, but with a photon counter. The same line is used in a
different QKD protocol: T. Debuisschert, W. Boucher, Phys. Rev. A
{\bf 70}, 042306 (2004).

\bibitem{unknown} H.-K. Lo, H.F. Chau, M. Ardehali, J. Cryptology {\bf 18},
133 (2005); see also quant-ph/9803007.



\bibitem{draft} N. Gisin et al., quant-ph/0411022

\bibitem{inoue} A similar argument applies to a different protocol: K. Inoue, T. Honjo, Phys.
Rev. A {\bf 71}, 042305 (2005)

\bibitem{decoy} W.-Y. Hwang, Phys. Rev. Lett. {\bf 91},
057901 (2003); X.-B. Wang, quant-ph/0410075; H.-K. Lo, X. Ma, K.
Chen, quant-ph/0411004.

\bibitem{modelocked} Alternatively, the source could be a pulsed
mode-locked laser followed by a pulse-picker.

\bibitem{noteopt} If dark counts can be neglected ($Q_{det}=0$), the
optimization can be done analytically: for the present protocol,
$\mu_{opt}\simeq V/[2(2-V-t)]$; for BB84,
$\mu_{opt}=f(V)/[2(1-t)]$ with, and $\mu_{opt}=tf(V)$ without
decoy states, where $f(V)=[V-h(\frac{1-V}{2})]$. For simplicity,
in the optimization we have taken $t_B\approx 1$ for both
protocols, although this may be technically harder to achieve in
BB84 because there are more optical components.

\bibitem{thew05} R. Thew et al., in preparation.

\end{thebibliography}
\end{document}